\documentstyle[aps,twocolumn]{revtex}
\begin{document}
\author{J. A. Jones\footnotemark[1] and E. Knill\footnotemark[2]}
\title{Efficient Refocussing of One Spin and Two Spin Interactions\\
for NMR Quantum Computation}
\address{\footnotemark[1]
Centre for Quantum Computation, Clarendon Laboratory, University of Oxford,
Parks Road, Oxford OX1 3PU, UK,\\ and Oxford Centre for Molecular Sciences,
South Parks Road, Oxford OX1 3QT, UK\\
\mbox{}\\
\footnotemark[2]Computer Research and Applications CIC-3, MS B-265, Los Alamos National Laboratory,
Los Alamos, New Mexico 87545}
\date{\today}
\maketitle

\begin{abstract}
The use of spin echoes to refocus one spin interactions (chemical
shifts) and two spin interactions (spin--spin couplings) plays a central
role in both conventional NMR experiments and NMR quantum computation.
Here we describe schemes for efficient refocussing of such interactions
in both fully and partially coupled spin systems.
\end{abstract}

\section{Introduction}
Much of the power and utility of NMR stems from the ease with which the
experimenter can control the effective Hamiltonian experienced by the
spin system. In conventional NMR \cite{Ernst} this permits different interactions to
be studied individually, while in NMR quantum computation
\cite{Cory96,Cory97,Gersh97,Jones98} this process is used to generate
Hamiltonians corresponding to quantum logic gates between specific spins
\cite{Jones98a}. This manipulation can be achieved using a variety of
techniques \cite{Freeman97}, but the simplest and most important
approach is the use of spin echoes. Applying a $180^\circ$ pulse to a
single spin in the middle of some evolution period acts to refocus any
evolution occuring as a result of one spin interactions (that is,
chemical shifts) or two spin interactions (spin--spin coupling,
assumed to be \emph{weak})
involving that spin. Thus the corresponding terms in the spin
Hamiltonian are effectively deleted.

This approach is simple to apply in systems containing only a small number
of coupled spins, but must be treated with caution when applied to
larger systems, especially when all the spins are coupled to one
another. The Hamiltonian describing evolution of a fully coupled spin
system of $N$ spins contains $N$ one spin terms and $N(N-1)/2$ two spin
terms, and it would be possible to produce a large number of effective
Hamiltonians, corresponding to any desired combination of these terms.
For simplicity we will concentrate on one particularly simple
Hamiltonian in which only one chemical shift term is retained; two
other simple Hamiltonians (in which only one coupling is retained, or
all interactions are refocussed) can be easily generated by small
modifications to the corresponding pulse sequences.

A pulse sequence for achieving this simplification in a two spin system
is shown in figure~\ref{fig:twospin}. In this case the refocussing
process is simple, and can be achieved with only two time periods,
separated by a single $180^\circ$ pulse.  For completeness it is
necessary to apply a final $180^\circ$ pulse to spin $1$ to ensure that
each spin experiences an even number of $180^\circ$ pulses ({\sc not}
gates), but in many cases such pulses can be omitted.  When applied to larger spin
systems, however, it is necessary to refocus many more interactions,
requiring a larger number of time periods. The conventional approach
\cite{Freeman99} is to recursively nest copies of sequences like
that shown in figure~\ref{fig:twospin} within one another, as shown for
a fully coupled four spin system in figure~\ref{fig:fourspin}. While
this nesting process is effective, it is exponentially inefficient, in
that the number of time periods and $180^\circ$ pulses required grows
exponentially with the number of spins in the spin system: while a four
spin system requires that the evolution period be divided into eight
sections, a five spin system will require sixteen sections, and so on.
\begin{figure}
\begin{center}
\begin{picture}(70,50)
\put(10,10){\line(1,0){50}}
\put(0,10){$I^1$}
\put(33,10){\framebox(4,12){}}
\put(60,10){\dashbox{2}(4,12){}}
\put(10,40){\line(1,0){50}}
\put(0,40){$I^0$}
\end{picture}
\end{center}
\caption{Pulse sequence for generating a simple effective Hamiltonian in
a two spin ($I^0$ and $I^1$) system; boxes correspond to $180^\circ$
pulses. This sequence generates the Hamiltonian corresponding to the
chemical shift of spin 0 ($I_z^0$); the final $180^\circ$ pulse, shown
as a dashed box, can often be omitted.}
\label{fig:twospin}
\end{figure}
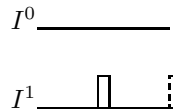
\begin{figure}
\begin{center}
\begin{picture}(220,110)
\multiput(10,10)(0,30){4}{\line(1,0){200}}
\multiput(33,10)(50,0){4}{\framebox(4,12){}}
\multiput(58,40)(100,0){2}{\framebox(4,12){}}
\put(108,70){\framebox(4,12){}}
\put(0,10){$I^3$}
\put(0,40){$I^2$}
\put(0,70){$I^1$}
\put(0,100){$I^0$}
\end{picture}
\end{center}
\caption{A pulse sequence for generating the effective Hamiltonian
$I_z^0$ in a fully coupled four spin system; boxes correspond to
$180^\circ$ pulses.}
\label{fig:fourspin}
\end{figure}
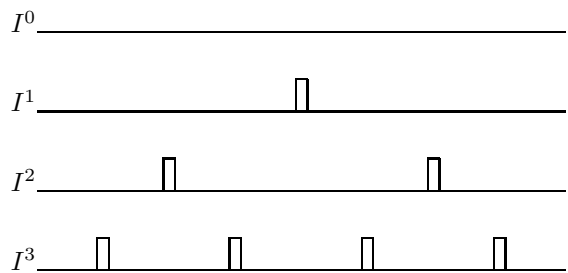

Although this nesting process is very widely used within NMR, a far more
efficient scheme is available. Here we describe this efficient
refocussing scheme, and show how it may be used to create the two
effective Hamiltonians described above\cite{Chuang}. We also discuss the
application of this scheme to partially coupled spin systems.

\section{Theory}
While spin echo sequences are usually drawn out as sequences of pulses,
as shown above, it is more convenient to describe a sequence
mathematically by considering the evolution of each spin during the
various equal time periods of free precession (the $180^\circ$ pulses are assumed
to be of negligible duration). Suppose that before the start of the spin
echo sequence a spin is in a state of $p=+1$ quantum coherence; the
effect of the $180^\circ$ pulse is to convert this to $-1$ quantum
coherence, and vice versa. The evolution of the spin will depend on its
coherence order, and thus the evolution during any period can be
described by a set of numbers taking the values $+1$ and $-1$, where the
value changes sign each time a $180^\circ$ pulse is applied to the spin.

These sets of values can then be gathered together into a matrix, whose
rows correspond to the individual spins and whose columns correspond to
the different time periods. Thus the sequence depicted in
figure~\ref{fig:twospin} can be described by the matrix
\begin{equation}
M_2=\left(\begin{array}{rr}
1 & 1 \\ 1 & -1
\end{array}\right),
\end{equation}
while that shown in figure~\ref{fig:fourspin} corresponds to the matrix
\begin{equation}
M_4=\left(\begin{array}{rrrrrrrr}
1 & 1 & 1 & 1 & 1 & 1 & 1 & 1\\
1 & 1 & 1 & 1 &-1 &-1 &-1 &-1\\
1 & 1 &-1 &-1 &-1 &-1 & 1 & 1\\
1 &-1 &-1 & 1 & 1 &-1 &-1 & 1
\end{array}\right).
\label{eq:Mfour}
\end{equation}
The refocussing effected by these pulse sequences can then be easily
explained by examining the properties of the corresponding matrices. The
chemical shift of a spin will be refocussed as long as the corresponding
row of the matrix $M$ contains the same number of plus and minus
ones. Similarly the coupling between two spins will be refocussed if the
vector obtained by multiplying pairs of numbers from the rows
corresponding to the two spins contains the same number of plus and
minus ones. More concisely, a spin--spin coupling will be refocussed if
the corresponding rows are orthogonal, while a chemical shift will be
refocussed if the corresponding row is orthogonal to a row of ones.

These two properties are sufficient to allow us to construct refocussing
matrices, and thus pulse sequences, with the desired properties.
Consider a system of $N$ spins, where we wish to refocus all the
interactions \emph{except} the chemical shift of spin $0$. This can be
achieved by constructing a refocussing matrix comprising one row of ones,
and $N-1$ rows of plus and minus ones, all of which are orthogonal to
one another and to the first row; the most efficient refocussing sequence
will correspond to the matrix with the smallest number of columns. Such
matrices are closely related to the well known Hadamard matrices.

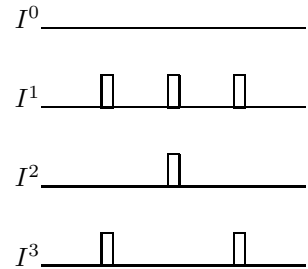
\begin{figure}
\begin{center}
\begin{picture}(120,110)
\multiput(10,10)(0,30){4}{\line(1,0){100}}
\multiput(33,10)(50,0){2}{\framebox(4,12){}}
\put(58,40){\framebox(4,12){}}
\multiput(33,70)(25,0){3}{\framebox(4,12){}}
\put(0,10){$I^3$}
\put(0,40){$I^2$}
\put(0,70){$I^1$}
\put(0,100){$I^0$}
\end{picture}
\end{center}
\caption{An efficient pulse sequence for generating the effective Hamiltonian $I_z^0$ in
a fully coupled four spin system.}
\label{fig:Hfour}
\end{figure}

The Hadamard matrix corresponding to a two spin system, $H_2$ takes the simple form
\begin{equation}
H_2=M_2=\left(\begin{array}{rr}
1 & 1 \\ 1 & -1
\end{array}\right),
\end{equation}
and so the conventional two spin sequence (figure~\ref{fig:twospin}) is, not surprisingly,
the most efficient.  The four spin Hadamard matrix, $H_4$, can be calculated using
\begin{equation}
H_4=H_2\otimes H_2=\left(\begin{array}{rrrr}
1 & 1 & 1 & 1\\
1 &-1 & 1 &-1\\
1 & 1 &-1 &-1\\
1 &-1 &-1 & 1
\end{array}\right),
\end{equation}
and is only half the size of its conventional equivalent, $M_4$
(equation~\ref{eq:Mfour}). The corresponding pulse sequence, shown in
figure~\ref{fig:Hfour}, is similarly shorter than the conventional
equivalent (figure~\ref{fig:fourspin}). Note that, with the exception of
spin $0$, there is no particular significance to the spin labels, and
individual nuclei can be assigned to the different spin numbers at will.

Unlike conventional spin echo sequences, these efficient sequences
involve the application of simultaneous $180^\circ$ pulses to two or
more spins. In principle this need not be a problem, but in practice it
may be necessary to choose $B_1$ field strengths with care, so as to
minimise the effects of Hartmann-Hahn transfers (whether homonuclear or
heteronuclear).

With slight modifications, these pulse sequences can also be used to
generate effective Hamiltonians corresponding to a single spin--spin
coupling. The procedure is straightforward: to generate a pure coupling
between spins $0$ and $1$ (for example), simply copy the pulses applied
to spin $1$ and apply them to spin $0$.  Similarly, a pulse
sequence in which \emph{all} one and two-spin
interactions are refocussed in an $N$ spin system
can be obtained by using the bottom $N$ lines of a sequence which refocusses
everything except the first chemical shift in an $N+1$ spin system.

The usefulness of this procedure depends on the existence of Hadamard
matrices with appropriate dimensions. Ideally the matrix should have the
same size as the number of spins in the spin system. Unfortunately it is
usually only possible to form Hadamard matrices whose dimension is a
multiple of four (the two by two Hadamard matrix, $H_2$, is a special
case); Hadamard matrices can be formed for many (though not all) multiples
of four, including all such multiples below fifty\cite{Griffiths,Hall}.
When the number of spins is \emph{not} a multiple of four, it is
instead necessary to use the next largest appropriate multiple of four, and select
an appropriate subset of the corresponding pulse sequence.  In general
this subset would probably be chosen to minimise the number of
simultaneous pulses in the pulse sequence. For example,
in a three spin system one possible pulse sequence is to use the lines
labelled $I^0$, $I^2$ and $I^3$ in figure~\ref{fig:Hfour}; this choice
corresponds to the conventional pulse sequence! This procedure is always
more efficient than the conventional approach except for the case of two
or three spins: in these cases the conventional and Hadamard approaches
give identical sequences.

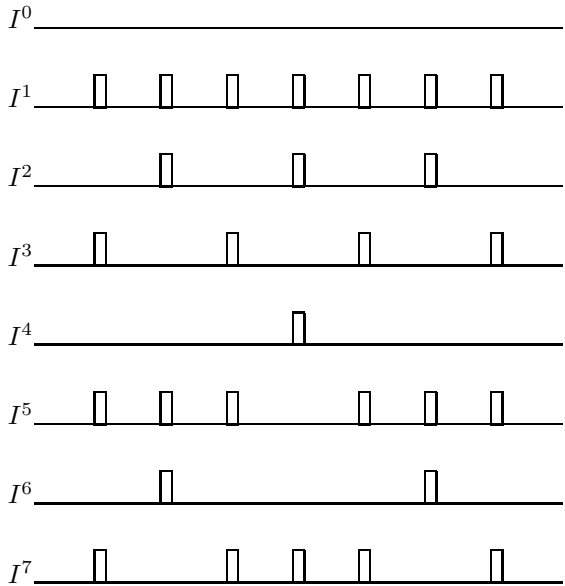
\begin{figure}
\begin{center}
\begin{picture}(220,230)
\multiput(10,10)(0,30){8}{\line(1,0){200}}
\multiput(33,10)(50,0){4}{\framebox(4,12){}}
\put(108,10){\framebox(4,12){}}
\multiput(58,40)(100,0){2}{\framebox(4,12){}}
\multiput(33,70)(25,0){3}{\framebox(4,12){}}
\multiput(133,70)(25,0){3}{\framebox(4,12){}}
\put(108,100){\framebox(4,12){}}
\multiput(33,130)(50,0){4}{\framebox(4,12){}}
\multiput(58,160)(50,0){3}{\framebox(4,12){}}
\multiput(33,190)(25,0){7}{\framebox(4,12){}}
\put(0,10){$I^7$}
\put(0,40){$I^6$}
\put(0,70){$I^5$}
\put(0,100){$I^4$}
\put(0,130){$I^3$}
\put(0,160){$I^2$}
\put(0,190){$I^1$}
\put(0,220){$I^0$}
\end{picture}
\end{center}
\caption{An efficient pulse sequence for generating the effective
Hamiltonian $I_z^0$ in a fully coupled eight spin system. The
conventional sequence would require 127 pulses distributed over 128
delay periods.}
\label{fig:Height}
\end{figure}

A discussed below, in real systems where only some of the possible couplings
are present it is not necessary to refocus all the couplings, and it is
instead possible to use simpler pulse sequences.  Spin coupling
networks in molecules are usually quite local, with resolved couplings
only being seen to a small number of close neighbours; it seems likely
that the four spin (figure~\ref{fig:Hfour}) and eight spin
(figure~\ref{fig:Height}) sequences would suffice for almost any system.

\section{Partially coupled spin systems}
In real life fully coupled spin systems are rather rare; in most cases
only a small subset of the possible couplings can be resolved. It is not
necessary to refocus all these unresolved couplings, which are
effectively absent, and thus the refocussing process can be greatly
simplified. In small spin systems it is practical to derive such
simplified sequences by hand, but in larger systems it is useful to have
an algorithmic procedure by which this can be achieved.

This is simply realised by treating the spin system as a non-complete
graph\cite{Biggs}. A graph comprises a set of vertices, connected by edges; this
corresponds to a set of nuclei connected by J-couplings. A partially
coupled spin system, in which some of the couplings are absent,
corresponds to an non-complete graph. A graph can be coloured, by
assigning each vertex one of a number of different colours, and the
colouring scheme is called a proper colouring if no two connected
vertices are the same colour. The graph may then be characterised by a
chromatic number, $\chi$: this is the smallest number of colours
required to properly colour the graph.  In a complete graph (a fully
coupled spin system) $\chi=N$, but in a partially coupled system $\chi$
can be much smaller.

The significance of this observation is that if a spin system is
represented by a properly coloured graph, then it is not necessary to
refocus interactions between nuclei corresponding to vertices with the
same colour.  To refocus all the interactions in an $N$ spin
system it suffices to create a pulse sequence corresponding to a $\chi$
spin system, and apply identical pulses to all nuclei with the same
colour.  In this case there is no need for further concern about
Hartman-Hahn transfers, as these additional simultaneous pulses will
only be applied to spins which are not coupled to one another.

Clearly this approach is only practical if it is easy to determine both
the value of $\chi$ and a corresponding proper colouring. In general
this is extremely difficult: indeed, determining $\chi$ is an NP hard
problem. This difficulty is, however, more apparent than real, as it is
relatively simple to estimate $\chi$, and to find corresponding proper
colourings, for certain simple types of graph such as those likely to
occur in coupled spin systems. If the maximum number of edges at any
vertex (that is, the maximum number of spins coupled to any other spin)
is $k$, then the graph is said to be of degree $k$, and $\chi\le k+1$;
in all but a few special cases $\chi\le k$. Furthermore, it is easy to
construct a proper colouring using at most $k$ (or $k+1$) colours.
Creating a sequence which refocuses all interactions except one chemical
shift or one J-coupling is slightly more complicated, but the simplest approach is to
assign the nuclei in question a unique colour; at worst this will
increase the number of colours required by 1.

\section{Conclusions}
The use of Hadamard matrices and non-complete graphs provides a powerful
language for describing pulse sequences which refocus one spin and two
spin interactions in NMR.  This approach permits the construction of
refocussing pulse sequences which are much shorter than their
conventional equivalents.

\section*{Acknowledgements}
We thank M.~Mosca for useful discussions. JAJ is a Royal Society
University Research Fellow.  EK thanks the National Security Agency for
financial support.


\begin{references}
\bibitem{Ernst} R.~R. Ernst, G.~Bodenhausen, and A.~Wokaun, ``Principles
of Nuclear Magnetic Resonance in One and Two Dimensions'' (Oxford University
Press, Oxford, 1987).

\bibitem{Cory96}
D.~G. Cory, A.~F. Fahmy, and T.~F. Havel, \emph{in} ``PhysComp '96''
(T.~Toffoli, M.~Biafore, and J.~Le\~{a}o, Eds.), pp. 87--91, New England Complex
Systems Institute (1996)

\bibitem{Cory97}  D. G. Cory, A. F. Fahmy and T. F. Havel, \emph{Proc. Nat. Acad.
Sc. USA} \textbf{94,} 1634 (1997).

\bibitem{Gersh97} 
N.~A. Gershenfeld and I.~L. Chuang, \emph{Science} \textbf{275,} 350 (1997).

\bibitem{Jones98} J. A. Jones and M. Mosca, \emph{J. Chem. Phys.} \textbf{109,}
1648 (1998).

\bibitem{Jones98a} J. A. Jones, R. H. Hansen, and M. Mosca, \emph{J. Magn. Reson.}
\textbf{135,} 353 (1998).

\bibitem{Freeman97}
R.~Freeman, ``Spin Choreography,'' Oxford University Press (1997).

\bibitem{Freeman99}
N.~Linden, H.~Barjat, R.~J. Carbajo, and R.~Freeman, \emph{Chem. Phys.
Lett.} (in press, see also quant-ph/9811043).

\bibitem{Chuang} A similar scheme has been derived indpendently by
D.~W. Leung \emph{et al.} (quant-ph/9904100).

\bibitem{Griffiths}
P.~R. Griffiths, ``Transform Techniques in Chemistry,'' Plenum Press,
New York (1978).

\bibitem{Hall}
M.~Hall~Jr, ``Combinatorial Theory,'' John Wiley and Sons, New York
(1986).

\bibitem{Biggs}
N.~L. Biggs, ``Discrete Mathematics,'' Clarendon Press, Oxford (1985).
\end{references}
\end{document}